\documentclass[apjl]{emulateapj}
\usepackage{amsmath}

\newcommand{\beginsupplement}{%
        \setcounter{table}{0}
        \renewcommand{\thetable}{S\arabic{table}}%
        \setcounter{figure}{0}
        \renewcommand{\thefigure}{S\arabic{figure}}%
     }

\shortauthors{Kreidberg et al.}

\begin{document}

\title{A precise water abundance measurement for the hot Jupiter WASP-43\lowercase{b}}

\author{
Laura Kreidberg\altaffilmark{1,12},
Jacob L. Bean\altaffilmark{1,13},
Jean-Michel D{\'e}sert\altaffilmark{2},
Michael R. Line\altaffilmark{3},
Jonathan J. Fortney\altaffilmark{3},
Nikku Madhusudhan\altaffilmark{4},
Kevin B. Stevenson\altaffilmark{1,14},
Adam P. Showman\altaffilmark{5},
David Charbonneau\altaffilmark{6},
Peter R. McCullough\altaffilmark{7},
Sara Seager\altaffilmark{8},
Adam Burrows\altaffilmark{9},
Gregory W. Henry\altaffilmark{10},
Michael Williamson\altaffilmark{10}, 
Tiffany Kataria\altaffilmark{5} \&
Derek Homeier\altaffilmark{11}
}

\email{E-mail: laura.kreidberg@uchicago.edu}

\altaffiltext{1}{Department of Astronomy and Astrophysics, University of Chicago, 5640 S.~Ellis Ave, Chicago, IL 60637, USA}
\altaffiltext{2}{CASA, Department of Astrophysical \& Planetary Sciences, University of Colorado, 389-UCB, Boulder, CO 80309, USA}
\altaffiltext{3}{Department of Astronomy and Astrophysics, University of California,Santa Cruz, CA 95064, USA}
\altaffiltext{4}{Institute for Astronomy, University of Cambridge, Cambridge CB3 OHA, UK}
\altaffiltext{5}{Department of Planetary Sciences and Lunar and Planetary Laboratory, The University of Arizona, Tuscon, AZ 85721, USA}
\altaffiltext{6}{Department of Astronomy, Harvard University, Cambridge, MA 02138, USA}
\altaffiltext{7}{Space Telescope Science Institute, Baltimore, MD 21218, USA}
\altaffiltext{8}{Department of Physics, Massachussetts Insitute of Technology, Cambridge, MA 02139, USA}
\altaffiltext{9}{Department of Astrophysical Sciences, Princeton University, Princeton, NJ 08544, USA}
\altaffiltext{10}{Center of Excellence in Information Systems, Tennessee State University, Nashville, TN 37209, USA}
\altaffiltext{11}{Centre de Recherche Astrophysique de Lyon, UMR 5574, CNRS, Universit\'e de Lyon, \'Ecole Normale Sup\'erieure de Lyon, 46 All\'ee d'Italie, F-69364 Lyon Cedex 07, France}
\altaffiltext{12}{National Science Foundation Graduate Research Fellow}
\altaffiltext{13}{Alfred P.~Sloan Research Fellow}
\altaffiltext{14}{NASA Sagan Fellow}

\begin{abstract}
The water abundance in a planetary atmosphere provides a key constraint on the planet's primordial origins because water ice is expected to play an important role in the core accretion model of planet formation.  However, the water content of the solar system giant planets is not well known because water is sequestered in clouds deep in their atmospheres.   By contrast, short-period exoplanets have such high temperatures that their atmospheres have water in the gas phase, making it possible to measure the water abundance for these objects.  We present a precise determination of the water abundance in the atmosphere of the 2\,$M_\mathrm{Jup}$ short-period exoplanet WASP-43b based on thermal emission and transmission spectroscopy measurements obtained with the \textit{Hubble Space Telescope}. We find the water content is consistent with the value expected in a solar composition gas at planetary temperatures ($0.4 - 3.5\times$ solar at 1\,$\sigma$ confidence).  The metallicity of WASP-43b's atmosphere suggested by this result extends the trend observed in the solar system of lower metal enrichment for higher planet masses.  
\end{abstract}


\section{INTRODUCTION}
Water ice is an important building block for planet formation under the core accretion paradigm \citep{pollack96}.  According to this model, protoplanetary cores form by sticky collisions of planetesimals.  Once the cores reach a threshold mass, they experience runaway accretion of nearby material.  Beyond the water frost line, water is expected to be the dominant component by mass of planetesimals in solar composition protoplanetary disks \citep{marboeuf08, johnson12}.  Measurements of a planet's water content can therefore help constrain the disk chemistry, location, and surface density of solids where it formed \citep[e.g.][]{lodders04, mousis09, oberg11, madhusudhan11, mousis12, helled14, marboeuf14}.

Despite water's important role in planet formation, there are few observational constraints on the bulk abundance of water in gas giant planets.  The solar system giants have such low temperatures that water has condensed into clouds deep in their atmospheres, and is not easily accessible to remote observations \citep{guillot14}.  The Galileo probe mass spectrometer entered Jupiter's atmosphere and provided a direct measurement of the water abundance, but found a surprisingly small value ($0.29\pm0.10\times$ solar) which contrasts with the $2-5\times$ solar enhancement of most other volatile species \citep{wong04}.  The reliability of the water measurement is uncertain given local meteorological effects at the probe entry point \citep{showman98} and it is generally considered a lower limit.  One of the main goals of NASA's JUNO mission, which is scheduled to arrive at Jupiter in 2016, is to make a new measurement of the atmospheric water abundance \citep{matousek07}.

In contrast to the solar system planets, hot exoplanets should harbor gaseous water in their observable atmospheres.  Detections of water have been reported for a number of giant exoplanets \citep{grillmair08,konopacky13,deming13,birkby13}, and some previous measurements have yielded precise constraints on the abundance of water in these objects \citep{lee13,line14, madhusudhan14}.  However, interpretation of past results has been challenging for cases when theoretical models do not provide good fits to the observed spectra \citep[e.g.][]{line14} and when measurement reproducibility has been questioned \citep[e.g.][]{swain09, gibson11}. 

The Wide Field Camera 3 (WFC3) instrument on the \textit{Hubble Space Telescope} (\textit{HST}) has enabled transit and eclipse observations of exoplanets that give repeatable results over year-long time baselines \citep{kreidberg14} and consistent measurements with multiple anaylsis techniques \citep{deming13, kreidberg14, knutson14, mccullough14}.  We use \textit{HST}/WFC3 to measure precise transmission and emission spectra for the 2 $M_\mathrm{Jup}$, short-period exoplanet WASP-43b that enable comparative planetology with gas giants in the solar system.

\section{Observations and Data Reduction}
We observed three full-orbit phase curves, three primary transits, and two secondary eclipses of WASP-43b with 61 \textit{HST} orbits as part of GO Program 13467.  During the observations, we obtained low-resolution time series spectroscopy with the WFC3 G141 grism over the wavelength range 1.1 to 1.7 $\mu$m.  The phase curves each span the entire orbital period of the planet (19.5 hours) and include coverage of a transit and eclipse, yielding a total of six transit and five eclipse observations.  Further details of the observing campaign are described in a companion paper \citep{stevenson14}.  We focus here on constraints on the planet's water abundance obtained from the transit and eclipse data.

\begin{figure*}
\resizebox{0.8\textwidth}{!}{\includegraphics{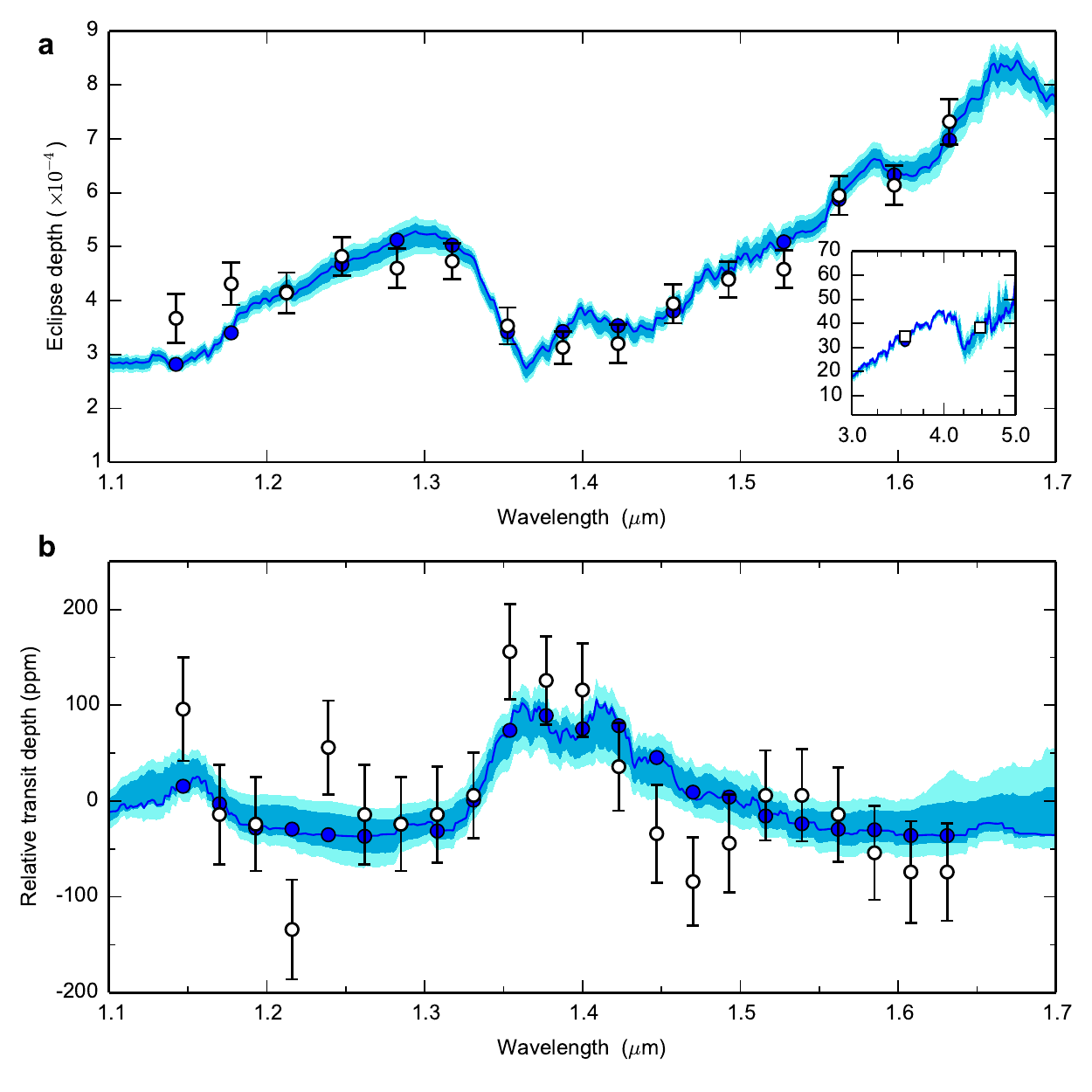}}
\centering
\caption{Emission and transmission spectra for WASP-43b. \textbf{a}, The emission spectrum measurements from \textit{HST}/WFC3 (white circles) and \textit{Spitzer}/IRAC (white squares; inset).   \textbf{b}, The transmission spectrum from WFC3 (white circles).  For both panels, the uncertainties correspond to 1\,$\sigma$ errors from a Markov chain fit.  The error bars for the \textit{Spitzer} measurements are smaller than the symbols.  We show the best fit models from our retrieval analysis (dark blue lines) with 1- and 2-$\sigma$ confidence regions denoted by blue and cyan shading.  The blue circles indicate the best fit model averaged over the bandpass of each spectroscopic channel.  The fits to both the emission and transmission spectra have chi-squared values nearly equal to the number of data points $n$ ($\chi^2$/$n$ = 1.2 for both).
}
\label{fig:spectra}
\end{figure*}

\begin{figure*}[h!]
\resizebox{1.0\textwidth}{!}{\includegraphics{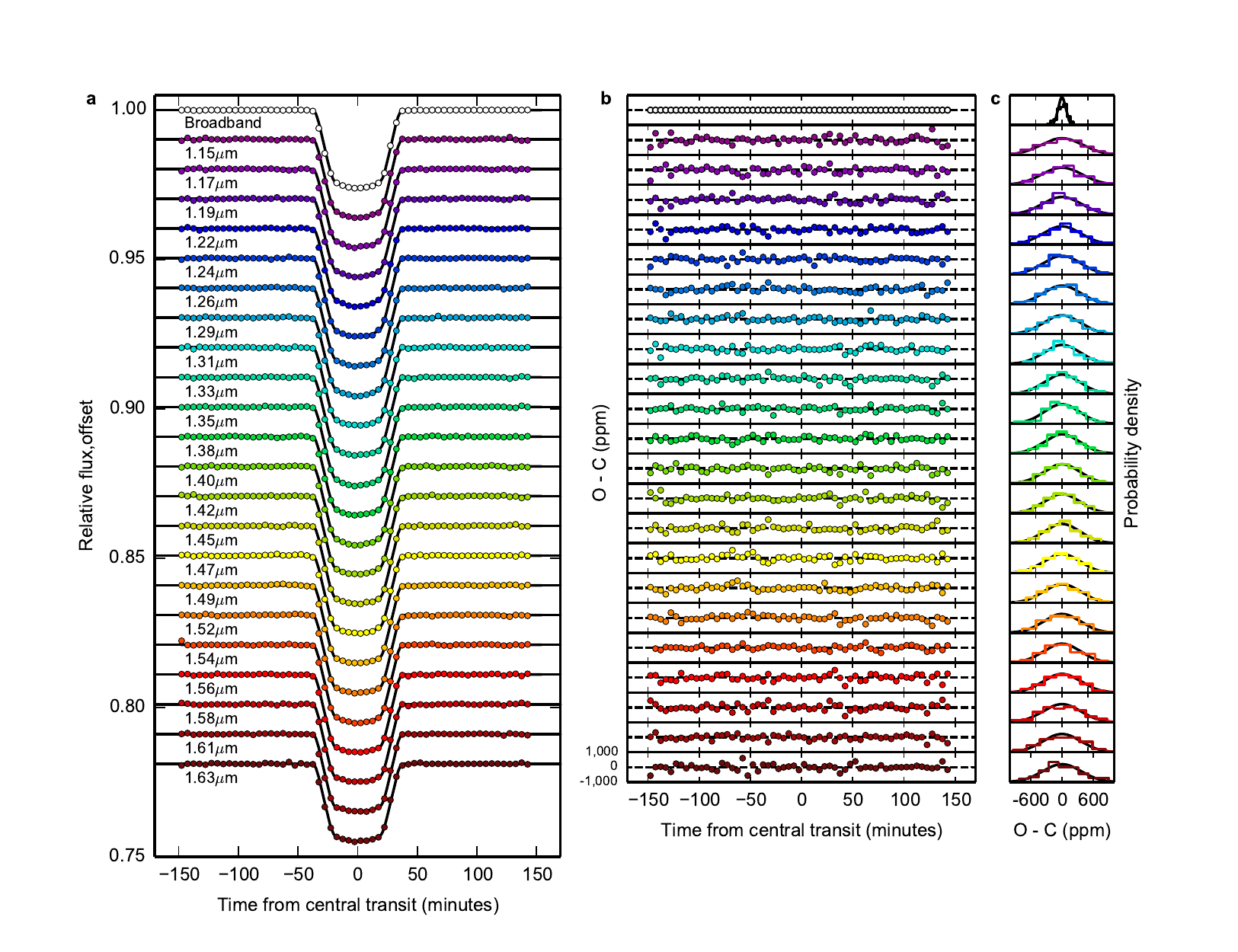}}
\caption{Spectrophotometric transit observations of WASP-43b. \textbf{a}, Transit light curves corrected for systematics (points) compared to the best-fit model light curves (lines).  The data are binned in 5-minute phase increments for visual clarity.  \textbf{b}, Residuals for the light curve fits, binned in phase.  The median residual root-mean-square (unbinned) is 353 parts per million, and all spectroscopic channels have residuals within 5\% of the photon noise limit. \textbf{c}, Histogram of the unbinned residuals (colored lines) compared to predictions from the photon noise (black lines).}
\label{fig:tlc}
\end{figure*}

\begin{figure*}[h!]
\resizebox{1.0\textwidth}{!}{\includegraphics{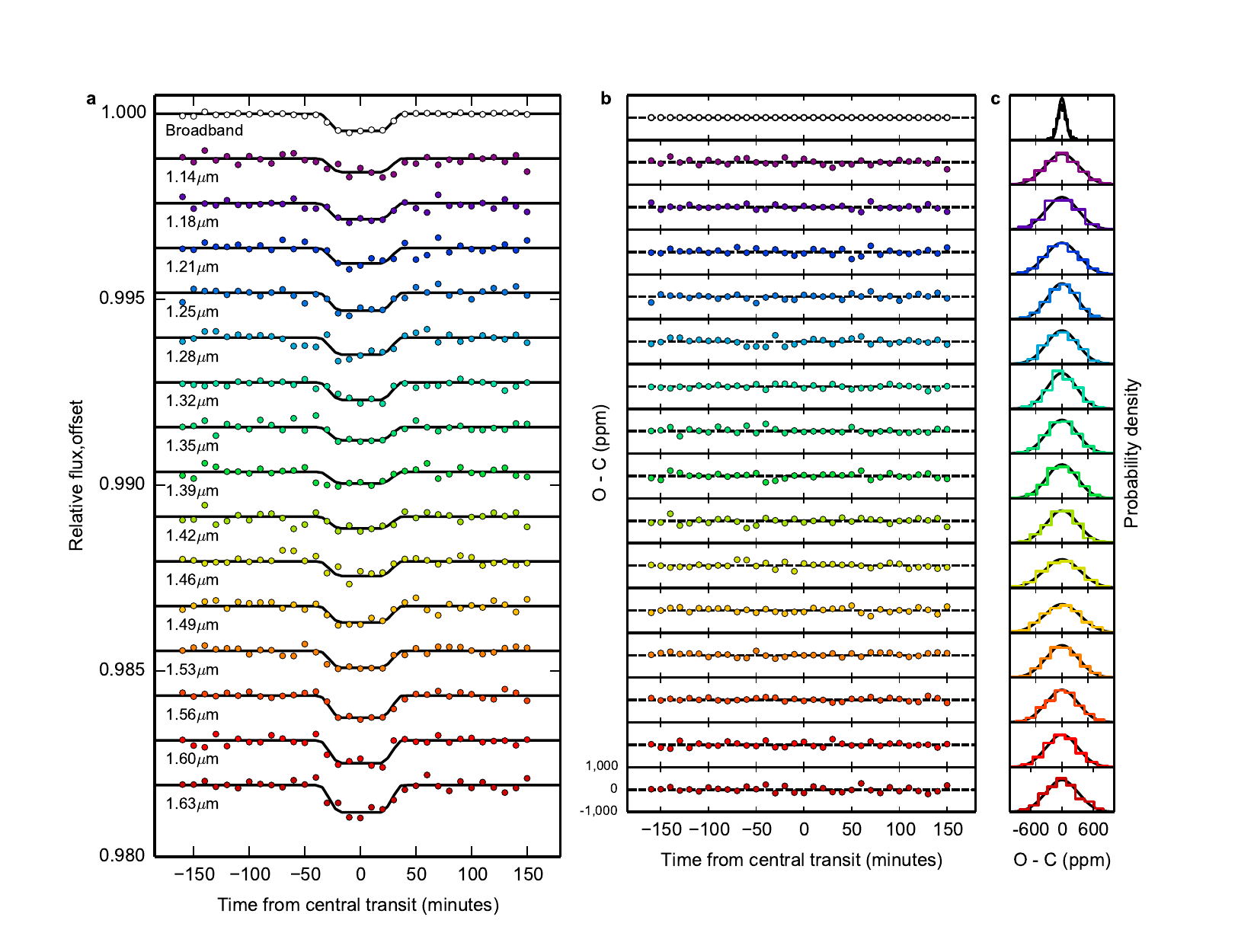}}
\caption{Eclipse observations of WASP-43b. \textbf{a}, Light curves for 15 spectrophotometric channels, binned in phase in 10-minute increments and corrected for instrument systematics (points).  The best-fit eclipse models are also shown (black lines).  \textbf{b}, Residuals for the light curve fits (also binned in phase).  \textbf{c}, Histograms of the unbinned residuals (colored lines) shown in comparison to the expected photon noise (black lines).}
\label{fig:elc}
\end{figure*}

We extracted spectroscopic light curves from the data using a technique outlined in past work \citep{kreidberg14}.  In our analysis, we used a subset of the total observations, including only spectra obtained within 160 minutes of the time of central transit or eclipse.  We fit the spectroscopic light curves to derive transmission and emission spectra, shown in Figure\,\ref{fig:spectra}.  The fitted light curves are shown in Figures \ref{fig:tlc} and \ref{fig:elc}.

The light curve fits consisted of either a transit or eclipse model \citep[as appropriate,][]{mandel02} multiplied by an analytic function of time used to correct systematic trends in the data.  The dominant systematic is an \textit{HST} orbit-long ramp \citep{berta12, deming13, kreidberg14, wilkins14}, which we fit with an exponential function \citep[using the \texttt{model-ramp} parameterization from][]{kreidberg14}.  The free parameters in our transit model are the planet-to-star radius ratio and a linear limb darkening coefficient.  The eclipse model has one free parameter, the planet-to-star flux ratio.  In all of our spectroscopic light curve fits, we fixed the orbital inclination to $82.1^\circ$, the ratio of semimajor axis to stellar radius to 4.872, and the time of central transit to 2456601.02748 BJD$_\mathrm{TDB}$ based on the best fit to the band-integrated (``white") transit light curve.  For the eclipse data, we also fixed the planet-to-star radius ratio to $R_p/R_s = 0.12$.  Our models use an orbital period equal to 0.81347436 days \citep{blecic14}.  The secondary eclipse time measured from the white eclipse light curve is consistent with a circular orbit, so we assume zero eccentricity for our spectroscopic light curve fits.

\begin{figure*}[h!]
\centering
\resizebox{0.6\textwidth}{!}{\includegraphics{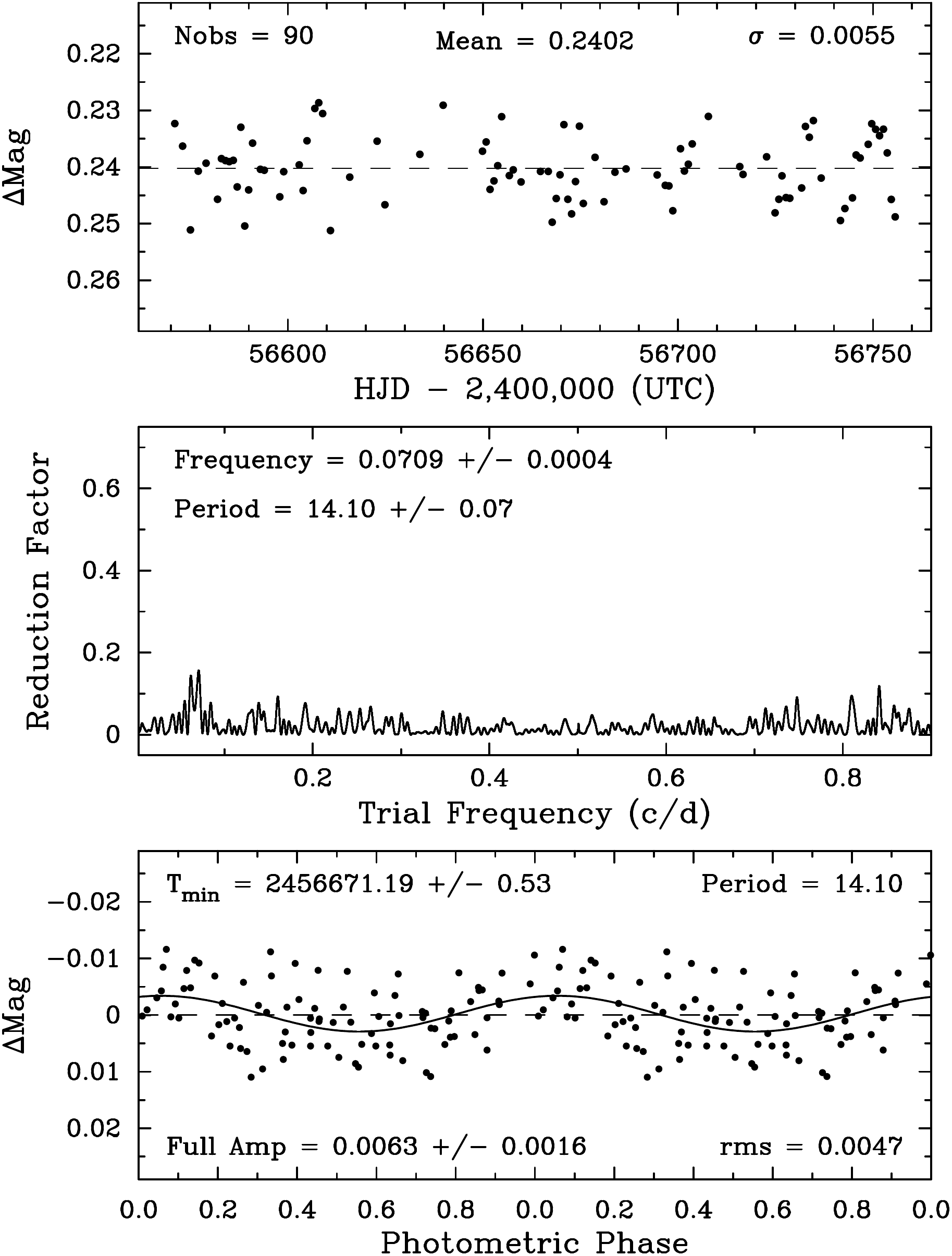}}
\caption{ (Top):  Cousins $R$ band photometry of WASP-43 during the 2013--2014 observing season acquired with the C14 automated imaging telescope at Fairborn Observatory.  Low amplitude brightness variability of approximately 0.005 mag is present. (Middle):  Frequency spectrum of the C14 observations, which suggests possible stellar rotation periods of 14.10 and 16.12 days.  (Bottom):  A least-squares sine fit of the C14 observations phased with the 14.10-day period shows reasonable coherence over the 2013--2014 observing season.}
\label{fig:phot}
\end{figure*}

We show the transit and eclipse depths from this analysis in Table\,\ref{tab:depths} and the transmission and emission spectra in Figure\,\ref{fig:spectra}.  All of the fitted light curves have residuals within 10\% of the predicted photon+read noise.  The median reduced chi-squared for the fits is 1.0 (for both transit and eclipse light curves). We measure consistent depths from epoch to epoch, which suggests that stellar variability does not significantly impact our measurements.  We obtained further confirmation of this from photometric monitoring of WASP-43 that shows minimal variation, indicating that the effect of starspots is below the precision of our data.  The photometry is presented in Figure \ref{fig:phot}.

\begin{deluxetable}{l r@{ $\pm$ } l  l c}
\tabletypesize{\scriptsize}
\tablecolumns{6}
\tablewidth{0pc}
\tablecaption{Transit and Eclipse Depths}
\tablehead{
 \colhead{Wavelength} & \multicolumn{2}{c}{Transit Depth\tablenotemark{a}} & Wavelength & Eclipse Depth \\
 \colhead{($\mu$m)} & \multicolumn{2}{c}{(ppm)} & \colhead{($\mu$m)} & \colhead{(ppm)}}
\startdata
1.135 -- 1.158 &  96 & 54 & 1.125 -- 1.160 & 367$ \pm $45\\ 
1.158 -- 1.181 & -14 & 52 & 1.160 -- 1.195 & 431$ \pm $39\\ 
1.181 -- 1.204 & -24 & 49 & 1.195 -- 1.230 & 414$ \pm $38\\ 
1.205 -- 1.228 & -134 & 52 & 1.230 -- 1.265 & 482$ \pm $36\\
1.228 -- 1.251 &  56 & 49 & 1.265 -- 1.300 & 460$ \pm $37\\ 
1.251 -- 1.274 & -14 & 52 & 1.300 -- 1.335 & 473$ \pm $33\\ 
1.274 -- 1.297 & -24 & 49 & 1.335 -- 1.370 & 353$ \pm $34\\ 
1.297 -- 1.320 & -14 & 50 & 1.370 -- 1.405 & 313$ \pm $30\\ 
1.320 -- 1.343 &   6 & 45 & 1.405 -- 1.440 & 320$ \pm $36\\ 
1.343 -- 1.366 & 156 & 50 & 1.440 -- 1.475 & 394$ \pm $36\\
1.366 -- 1.389 & 126 & 46 & 1.475 -- 1.510 & 439$ \pm $33\\ 
1.389 -- 1.412 & 116 & 49 & 1.510 -- 1.545 & 458$ \pm $35\\ 
1.412 -- 1.435 &  36 & 46 & 1.545 -- 1.580 & 595$ \pm $36\\ 
1.435 -- 1.458 & -34 & 51 & 1.580 -- 1.615 & 614$ \pm $37\\ 
1.458 -- 1.481 & -84 & 46 & 1.615 -- 1.650 & 732$ \pm $42\\ 
1.481 -- 1.504 & -44 & 51 &&\\
1.504 -- 1.527 &   6 & 47 &&\\ 
1.527 -- 1.550 &   6 & 48 &&\\ 
1.550 -- 1.573 & -14 & 49 &&\\ 
1.573 -- 1.596 & -54 & 49 &&\\ 
1.596 -- 1.619 & -74 & 53 &&\\
1.619 -- 1.642 & -74 & 51 &&
\enddata
\tablenotetext{a}{Transit depths are given relative to the mean over all wavelengths, which is 2.5434\%.}
\label{tab:depths}
\end{deluxetable}

\section{Analysis}
We retrieved the planet's atmospheric properties by fitting the transmission and dayside emission spectra with the CHIMERA Bayesian retrieval suite \citep{line13a,line13b,line14}.  The retrieval constrains the molecular abundances and the temperature-pressure (T-P) profile of WASP-43b's atmosphere.  In addition to our \textit{HST} data, we included two high-precision, 3.6- and 4.5-$\mu$m broadband photometric \textit{Spitzer Space Telescope}/IRAC secondary eclipse measurements \citep{blecic14} in the retrieval.  We found that using the highest-precision ground-based secondary eclipse measurements \citep{gillon12} did not significantly affect our results, so our final analysis incorporates data from \textit{HST} and \textit{Spitzer} only. 

We analyzed the transmission and emission spectra independently.  In both cases, we retrieved the abundances of H$_2$O, CH$_4$, CO, and CO$_2$, which are expected to be the dominant opacity sources at the observed wavelengths for a hydrogen-rich atmosphere.  Our model also includes collision-induced H$_2$/He absorption.  We explored the effects of including the additional chemical species H$_2$S, NH$_3$, K, and FeH, and found that our results were unchanged.  

\begin{figure*}
\resizebox{1.0\textwidth}{!}{\includegraphics{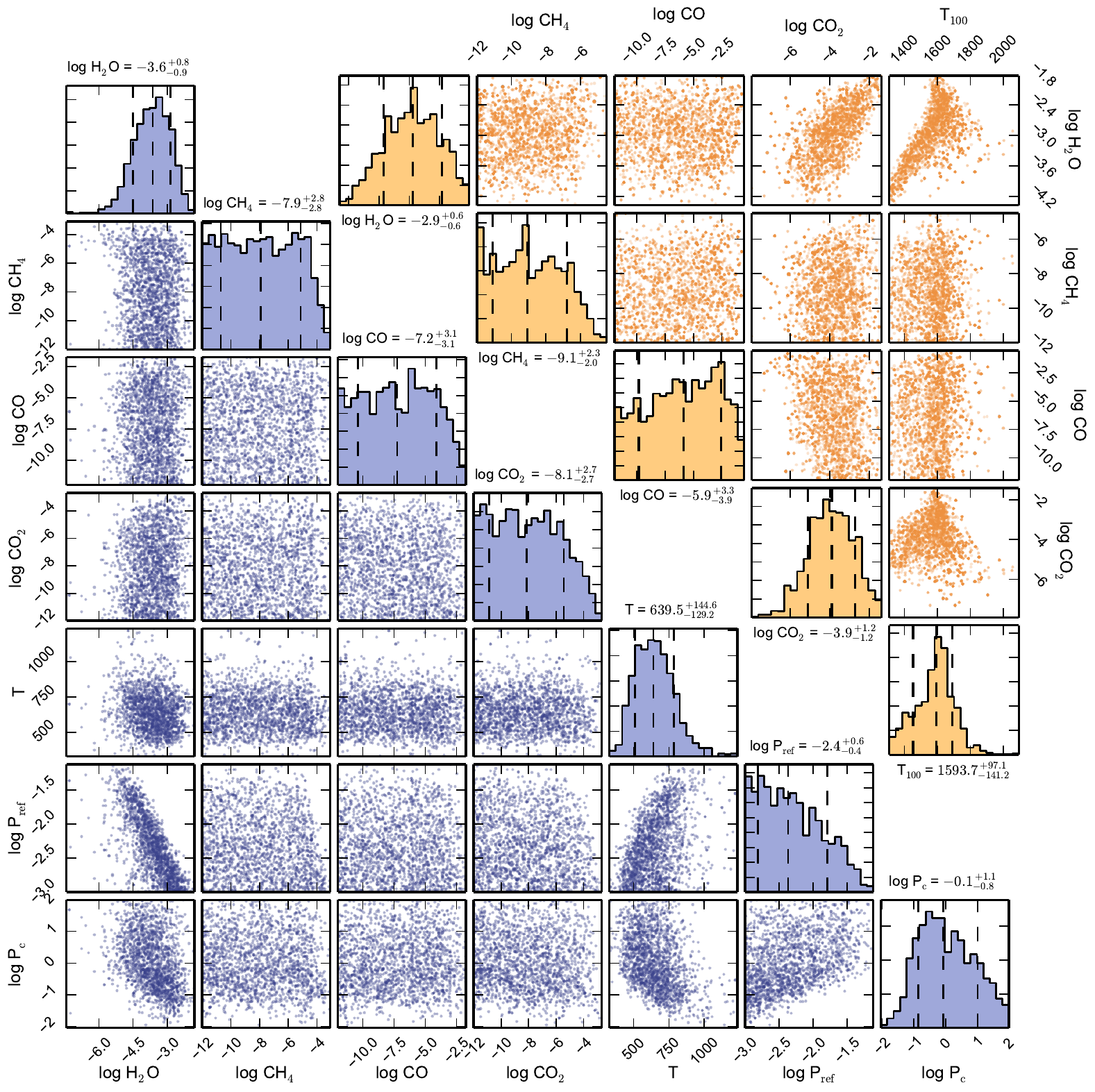}}
\caption{Pairs plots of retrieved parameters from the emission spectrum (top right) and the transmission spectrum (lower left).  We plot every tenth point from our MCMC chains.  For the emission spectrum fits, we show constraints on the retrieved molecular abundances (in units of log mixing ratio) and the temperature at the 100 mbar pressure level ($T_{100}$, in Kelvin). For the transmission spectrum, we show constraints on the molecular abundances (in units of log mixing ratio), the scale height temperature (in Kelvin), the reference pressure $P_\mathrm{ref}$ (in bars), and the cloud-top pressure $P_c$ (in bars).}
\label{fig:pairs}
\end{figure*}

For the emission spectrum retrieval, we used a five-parameter model for the T-P profile motivated by analytic gray radiative-equilibrium solutions \citep{parmentier14}.  The model fits a one-dimensional T-P profile to the hemispherically averaged emission spectrum.  For the transmission spectrum modeling, we retrieved an effective scale height temperature, a reference pressure at which the fiducial radius is defined, and an opaque gray cloud top pressure, in addition to the molecular abundances \citep{line13c}.  The distributions of retrieved parameters are shown in Figure\,\ref{fig:pairs}.  The retrieved T-P profile for the dayside emission spectrum is presented in \cite{stevenson14}.

\section{Results}
\subsection{Constraints from the Emission Spectrum}
The emission spectrum shows strong evidence for water absorption.  We detect water at $11.7\,\sigma$ confidence ($6.4\,\sigma$ from the WFC3 data alone), according to the Bayesian information criterion (BIC).  The data constrain the volume mixing ratio of H$_2$O in the planet's atmosphere to be $3.1\times10^{-4}$ -- $4.4\times10^{-3}$ at $1\,\sigma$ confidence.  Figure\,\ref{fig:waterconstraint} shows the distribution of H$_2$O abundances that fit the observations.  

Water is the only molecule significantly detected over the WFC3 wavelength range; however, additional constraints from the \textit{Spitzer} data suggest CO and/or CO$_2$ are also present in the planet's atmosphere.  We detect CO+CO$_2$ at $4.7\,\sigma$ confidence in the combined WFC3/IRAC spectrum.  The measured abundance of CO$_2$ is relatively high compared to the expected thermochemical value for a solar composition gas ($\sim5\times10^{-8}$).  We find that a moderately super-solar metallicity composition has equilibrium H$_2$O, CO$_2$, and CO abundances that are within the range of our retrieved values.  However, the CO+CO$_2$ constraints are driven mainly by the photometric point from the \textit{Spitzer} 4.5 $\mu$m channel. The bandpass for this channel is about 1 $\mu$m wide, and covers features from CO, CO$_2$, and H$_2$O \citep{sharp07}.   Making a robust determination of the abundances of these molecules requires spectroscopic observations to resolve their absorption features.  The main conclusions of this work are unchanged if we exclude the \textit{Spitzer} data from our analysis.  

The best-fit thermal profile has decreasing temperature with pressure and is consistent with predictions from a radiative-convective model for the substellar point over the range of pressures to which our data are sensitive.  We find no evidence for a thermal inversion.  Further details of the thermal structure of the planet's atmosphere are available in \cite{stevenson14}.

\subsection{Constraints from the Transmission Spectrum}
We obtain complementary results for the atmospheric composition based on a retrieval for the transmission spectrum.  Water absorption is detected at 5\,$\sigma$ confidence and is visible in the data, shown in Figure\,\ref{fig:spectra} (panel b).  The transmission spectrum fit allows a water volume mixing ratio between 3.3$\times10^{-5}$ and 1.4$\times10^{-3}$ at 1\,$\sigma$, which is consistent with the bounds derived from the emission measurements.  No other molecules are detected in the spectrum according to the BIC.  The constraints on the water abundance (shown in Figure\,\ref{fig:waterconstraint}) are broader than those obtained from the emission spectrum because the abundance is correlated with the reference pressure level, which is only weakly constrained by the observations.  We remind the reader that the size of features in the transmission spectrum is controlled by the molecular abundances, the planet's atmospheric scale height, and the radius of the planet relative to the star \citep{millerricci09}.

In addition to probing the atmospheric composition, the data also constrain the temperature at the terminator and the cloud top pressure.  The 1\,$\sigma$ confidence interval on the scale height temperature is 500--780\,K.  We find no evidence for a cloud at the pressure levels to which our observations are sensitive.  

\subsection{Joint Constraint on the Water Abundance}
\begin{figure}
\resizebox{0.45\textwidth}{!}{\includegraphics{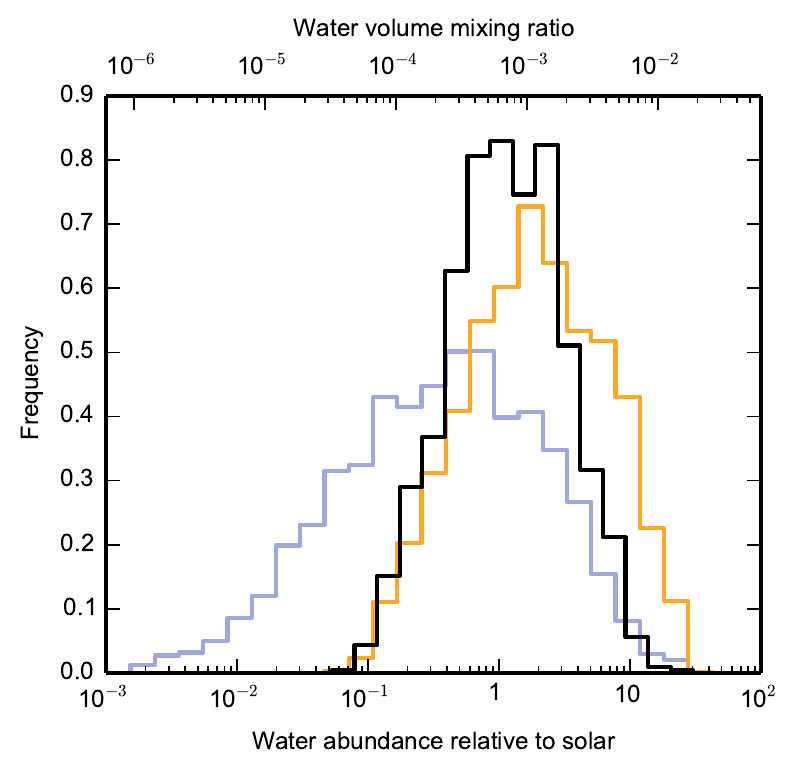}}
\caption{Constraints on the water abundance of WASP-43b. We show the probability distributions of water abundances measured from the emission spectrum (orange line) and transmission spectrum (blue line).  The water abundance relative to solar is calculated using a ``solar water abundance" of $6.1\times10^{-4}$, which is the water volume mixing ratio for a solar composition gas in thermochemical equilibrium at planetary temperatures.  The constraints from the emission and transmission spectra are consistent with each other, in accordance with expectations from theoretical models that the water abundance is constant to within a factor of $\sim2$ over the pressure levels and temperatures in the observable atmosphere \citep{moses11}.  The joint constraint from multiplying the two probability distributions is indicated by the solid black line, and constrains the water abundance to between $3.1\times10^{-4}$ and $4.4\times10^{-3}$ ($0.4 - 3.5\times$ solar) at $1\,\sigma$ confidence.}
\label{fig:waterconstraint}
\end{figure}

We derive consistent water abundances for WASP-43b from the emission and transmission spectra. This consistency matches the prediction from theoretical models of hot Jupiters that water has a nearly uniform abundance with both pressure (from $10$ to $10^{-8}$ bar) and with temperature \citep{moses11}.  Therefore, to obtain a more precise estimate of WASP-43b's water abundance, we assume the regions of the atmosphere probed by the emission and transmission data have the same water content.  Because the measurements are independent, we can combine their constraints by multiplying the probability distributions for water abundance retrieved from each data set.  This yields a joint distribution, shown in Figure\,\ref{fig:waterconstraint}, which constrains the water volume mixing ratio to $2.4\times10^{-4}$ -- $2.1\times10^{-3}$ at $1\,\sigma$ confidence.  This measurement contrasts with the sub-solar water abundance values reported for three other hot Jupiters by \cite{madhusudhan14}, and suggests that additional observations are needed to understand the diversity in composition of these objects.

\section{Discussion}
\subsection{Comparison with Solar System Planets}
With our well-determined water abundance for the atmosphere of WASP-43b, we can begin a comparative study with the giant planets in the solar system.  However, it is not possible to directly compare water abundances, because the water content in the solar system giants is poorly constrained.  We instead compare the planets' metallicities, which we estimate from chemical species with well-determined abundances.  To calculate metallicity based on a molecule $X$, we determine the planet's enhancement in $X$ relative to the volume mixing ratio of $X$ expected for a solar composition gas at planetary temperatures.  We use solar abundances from \cite{asplund09} for the calculation. 

\begin{figure}
\resizebox{0.5\textwidth}{!}{\includegraphics{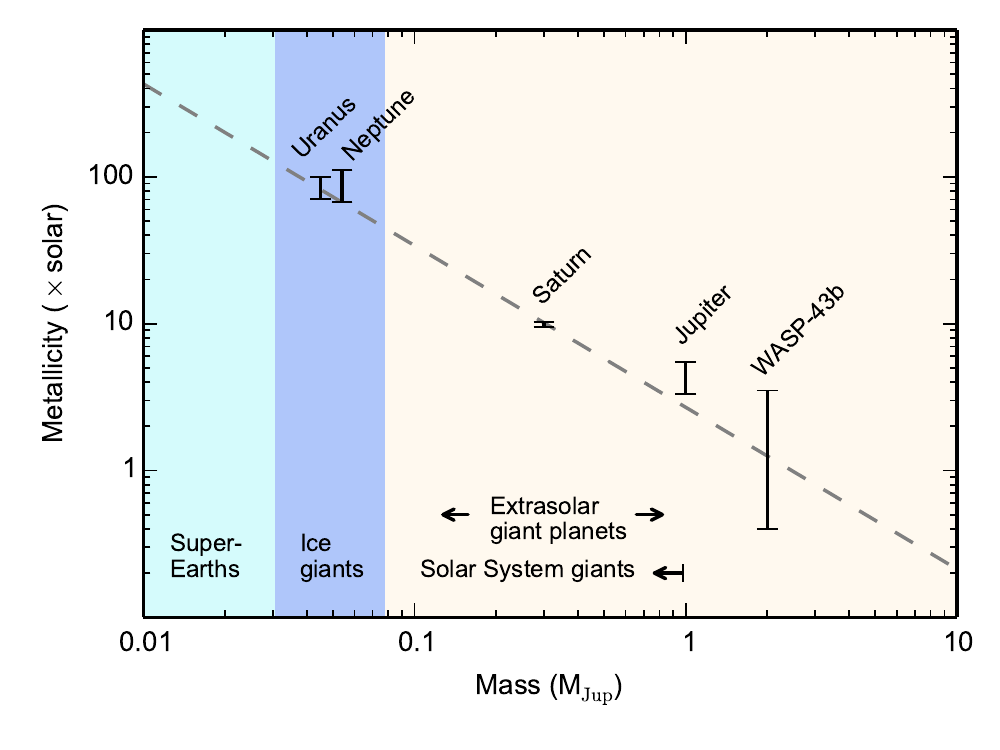}}
\caption{Atmospheric metal abundances as a function of planet mass for the solar system giant planets and WASP-43b.  The error bars correspond to 1\,$\sigma$ uncertainties and represent statistical errors only. The dashed line is a power law fit to the data.  We base the metallicity estimates on the abundances of representative proxies.  The solar system planet metallicities are inferred from methane, and the metallicity of WASP-43b is based on our derived water abundance.}  
\label{fig:Mtrend}
\end{figure}

We infer the solar system planets' metallicities from the abundance of methane, which has been precisely measured for all four giant planets.  Jupiter's methane abundance is from the Galileo probe \citep{wong04}, while those of the other planets are from infrared spectroscopy \citep{fletcher09a, karkoschka11, sromovsky11}.  The 1\,$\sigma$ confidence intervals for the planets' metallicities are 3.3--5.5, 9.5--10.3, 71--100, and 67--111\,$\times$ solar for Jupiter, Saturn, Neptune, and Uranus, respectively.  These bounds are shown in Figure\,\ref{fig:Mtrend}.

We determine the metallicity of WASP-43b based on our measured water abundance.  The planet's temperature is cooler at the terminator than at the substellar point. The difference between these temperatures leads to a factor of two discrepancy in the expected water volume mixing ratio for a solar composition gas.  This difference is small relative to the uncertainty in the planet's measured water abundance, so we therefore adopt the average predicted water volume mixing ratio ($6.1\times10^{-4}$).  Using the joint constraint from the transmission and emission spectra, we find the water abundance is $0.4$ to $3.5\times$ solar at $1\,\sigma$ confidence. The 3\,$\sigma$ upper limit on the water enhancement is 20$\times$ solar.

We note that determining the metallicity relative to solar composition assumes that the planets have a scaled solar abundance pattern.  This assumption could lead to an incorrect estimate of metallicity in the case of non-solar abundance ratios.  For example, if WASP-43b had a super-solar carbon-to-oxygen (C/O) ratio, we would expect a smaller fraction of the total oxygen to be partitioned into H$_2$O \citep{madhusudhan12}. We would therefore underestimate the planet's metallicity based on our assumption of solar abundances.  However, given that the C/O ratio is poorly constrained by our data but broadly consistent with solar, we proceed with the comparison with these caveats in mind.

The metallicity estimates for the solar system planets show a pattern of decreasing metal enhancement with increasing planet mass (see Figure\,\ref{fig:Mtrend}).  This trend is generally thought to be controlled by the relative importance of accretion of solid planetesimals versus H/He-dominated gas.  Planet population synthesis models aim to match atmospheric metallicity to planet mass \citep{fortney13,marboeuf14}, but there are limited data available for planets outside the solar system.  The metallicity of WASP-43b indicates that the trend seen in the solar system may extend to exoplanets.

\subsection{Prospects for Future Work}
WASP-43b, with twice the mass of Jupiter and an orbital period of less than one day, exemplifies the opportunity exoplanets provide to study planet formation over a larger parameter space than what is available in our solar system.  A more insightful comparative planetology study using WASP-43b could be performed by improving the precision of the water abundance estimate and measuring the abundances of additional molecules. Such measurements will be enabled by the broad wavelength coverage and increased sensitivity of next-generation observing facilities such as the \textit{James Webb Space Telescope}. 

However, a planet's chemical composition depends on many factors, including the planet's formation location within the protoplanetary disk, the composition, size and accretion rate of planetesimals, and the planet's migration history. Even perfect constraints on the abundances of many chemical species for a small number of objects may not yield a unique model for the origin of giant planets. Fortunately, the plethora of transiting exoplanets that have already been found and will be discovered with future missions offer the potential for statistical studies. Measuring precise chemical abundances for a large and diverse sample of these objects would facilitate the development of a more comprehensive theory of planet formation.

\acknowledgments
This work is based on observations made with the NASA/ESA Hubble Space Telescope that were obtained at the Space Telescope Science Institute, which is operated by the Association of Universities for Research in Astronomy, Inc., under NASA contract NAS 5-26555. These observations are associated with program GO-13467. Support for this work was provided by NASA through a grant from the Space Telescope Science Institute, the National Science Foundation through a Graduate Research Fellowship (to L.K.), the Alfred P. Sloan Foundation through a Sloan Research Fellowship (to J.L.B.), and the Sagan Fellowship Program (to K.B.S.) as supported by NASA and administered by the NASA Exoplanet Science Institute. G.W.H. and M.W.W. acknowledge long-term support from NASA, NSF, Tennessee State University, and the State of Tennessee through its Centers of Excellence program. D.H. acknowledges support from the European Research Council under the European Community's Seventh Framework Programme (FP7/2007-2013 Grant Agreement no. 247060).

\bibliographystyle{apj}
\bibliography{ms.bib}

\clearpage
\beginsupplement

\begin{center}
\LARGE{\textbf{Supplemental Figures}}
\end{center}

\setlength{\abovecaptionskip}{25pt plus 3pt minus 2pt}

\end{document}